# A Pharmacy Benefit Manager Insurance Business Model

Lawrence W. Abrams, Ph.D.

March 10, 2025

**Summary:**

It is time to move on from attempts to make the pharmacy benefit manager (PBM) reseller business model more transparent. Time and time again the Big 3 PBMs have developed opaque alternatives to piece-meal 100% pass-through mandates. Time and time again PBMs have demonstrated expertise in finding loopholes in state government disclosure laws.

The purpose of this paper is to provide quantitative estimates of two transparent insurance business models as a solution to the PBM agency issue. The key parameter used is an 8% gross profit margin figure disclosed by the Big 3 PBMs themselves. Based on reported drug trend delivered to plans, we use a $1,200 to $1,500 per member per year (PMPY) as the range for this key performance indicator (KPI).

We propose that discussions of PBM insurance business models start with the following figures: (1) a fixed premium model with medical loss ratio ranging from 92% to 85%; (2) a fee-for-service model ranging from $96 to $180 PMPY with risk sharing of deviations from a contracted PMPY delivered drug spend.

**Disclosures:**

I have not received any remuneration for this paper nor financial interest in any company cited in this working paper. I have a Ph.D. in Economics from Washington University in St. Louis and a B.A. in Economics from Amherst College. Other papers on PBMs can be found on my website  https://labrams.co

**The Opaque PBM Reseller Business Model**

The Big 3 pharmacy benefit managers (PBMs) -- CVS Caremark, Aetna Express Scripts, and United Healthcare OptumRx -- have been under attack for not acting in the best interest of plan sponsors due to an opaque reseller business model. Our contribution to the PBM debate began in 2003 when we first disaggregated Express Scripts' 10-Q financials showing retained formulary rebates exceeding 30%.[1] Our disaggregation of the PBM Medco's business model revealed a dramatic shift after 2005 to mail order generic margins.[2]

The growth of the PBM mail order business was enabled by a collusive hold-up of retail generic margins in order to show plan sponsors that PBM owned mail order pharmacies were a lower cost alternative. Walmart's loss-leader cash-only $4 / generic prescription (Rx) proved to be the first blow to this hold-up scheme. The 2006 vertical merger of retailer CVS and the PBM Caremark created the first incentive for a PBM to roll out a preferred provider retail pharmacy network. This marked the end of large percentage margins for generic prescriptions and the beginning of the financial decline of retail drugstores.[3] The next phase of the PBM reseller business model once again became dependent on retained formulary rebates, but this time it was rebates from self-injectable biologic drugs.[4]

______________________

1 Lawrence W. Abrams, *Estimating the Rebate Retention Rate of Pharmacy Benefit Managers,* April, 2003, https://c1c0481a-8d34-49dc-ad66-d6c488c905a9.usrfiles.com/ugd/c1c048_a05060d8eead4427bc95ab12e9815e83.pdf
2 Lawrence W. Abrams, *Pharmacy Benefit Managers as Conflicted Countervailing Powers,* January, 2007, https://c1c0481a-8d34-49dc-ad66-d6c488c905a9.usrfiles.com/ugd/c1c048_2b204a6edde64f22a60fb706717ddfb7.pdf
3 Lawrence W. Abrams, *CVS - Caremark and the Coming Preferred Provider Wars,* December 2006, https://c1c0481a-8d34-49dc-ad66-d6c488c905a9.usrfiles.com/ugd/c1c048_29cb633a83ef4ce4a797c01402546dd6.pdf
4 Lawrence W. Abrams, *Three Phases of the Pharmacy Benefit Manager Business Model,* September 2017, https://c1c0481a-8d34-49dc-ad66-d6c488c905a9.usrfiles.com/ugd/c1c048_e24040fcf1774f71b4438018d1f6d29f.pdf

Starting around 2012, under pressure from plan sponsors to pass through more formulary rebates and hold down list price inflation, PBMs added opaque administrative fees and price protection rebates to formulary bid menus. We are aware of only Nephron Research who has attempted to sort out the myriad ways PBMs currently profit from formulary management. Nephron Research data indicates that administrative fees, price protection rebates and mark-ups on specialty drugs now outweigh traditional retrained formulary rebates as a source of PBM gross profits.[5]

Critics of Big 3 PBMs finally need to realize that PBMs have a proven ability to maintain long term gross profit margins in the face of piece-meal 100% pass-through mandates. Similarly, PBMs can easily find loopholes in state government transparency laws by reclassifying suspect line items as something else. A recent analysis by Drug Channels highlights the futility of state transparency laws by pointing out the absurdity of some of the state government findings.[6]

**An Outline of a PBM Insurance Business Model**

The purpose of this section is to provide quantitative estimates of key financial parameters and key performance indicators (KPIs) for two PBM insurance business models: (1) a capitated premium model tied to a medical loss ratio (MLR); (2) a fee-for-service (FFS) model that includes a KPI of a per member per year (PMPY) delivered drug trend with risk sharing for deviations.

___________________

We think that our estimates could be a starting place for discussions between plan sponsors and PBMs around drug benefit plans underwritten by an insurance business model. We propose that the starting point for MLR and FFS discussions be the Big 3 PBM self-disclosed 8% long term average gross profit margin.[7]

The Big 3 PBMs yearly publish a Drug Trend Report highlighting new utilization management techniques, past year's outcomes, and future years' projections. Trends are measured as year-over-year percentages. While transparent, percentage KPIs are problematic when comparing different PBMs. The Big 3 PBMs stopped reporting average dollar PMPY drug trend after 2017. In that year, CVS Caremark and Express Scripts reported an average of $1,067 PMPY and $1,100 PMPY, respectively.[8]

Our search of small transparent PBM websites found only Navitus reporting a trend of $1,221 PMPY.[9] The State of California annually collects PMPY trend figures from large commercial plans licensed by the state.[10] In 2023, they disclosed that plans self-reported an average of $945 PMPY. Plans also self-reported a calculated MLR ranging from 90% to 77%. None of these self-reported figures were audited.

In our opinion, the reluctance of both big and small PBMs to publish their PMPY delivered trend is a clear indicator of an industry-wide reluctance to compete on this KPI.

________________

The figure used to derive key parameters for our insurance business models is an 8% long term average gross profit margin reported by the Big 3 PBMs themselves to hired consultants.[11] It was disclosed by the consultants in a recent 126-page rebuttal to an FTC administrative complaint of unfair competition.

It is interesting to note that the Big 3 PBMs also disclosed an average sale general and administrative expense (SG&A) ratio of 3% of net sales over the period. Healthcare insurance companies report SG&A ratios that are considerably higher due to the relative complexity of managing a medical benefit as opposed to a pharmacy benefit.

### Recasting the PBM Reseller Business Model

| | | | | | |
|---|---|---|---|---|---|
| 2017-2022 Big 3 PBM gross profit margin | 8% | | | | |
| Medical loss ratio (MLR) equivalent | 92% | | | | |
| | | | | | |
| Range of negotiated MLR | | High> | 92% | Low > | 85% |
| Range of margins for FFS equivalents | | Low > | 8% | High> | 15% |
|    based on $1,200 PMPY delivered trend | | Low > | $96 | High> | $180 |
| | | | | | |
| Range of negotiated delivered trend | | $1,200 to $1,500 PMPY | | | |

________________

11 Carlton, et. a*l., supra*, 79-80.

Plan sponsors' total drug benefit cost is the sum of delivered trend -- in the $1,200 to $1,500 PMPY range -- plus the cost of management -- in the $96 to $180 PMPY range.

These figures highlight the fact that a PBM delivered drug trend is more than ten times any PBM gross profit expressed as a FFS. PBM critics rarely discuss the implications of this ratio. Despite PBM misalignment, the size of the Big 3 PBMs allows them to negotiate greater formulary rebates from Pharma, likely resulting in an overall lower total benefit cost than a small transparent PBM. This is despite the latter offering a lower transparent FFS.

From the standpoint of compensating PBMs on a risk-adjusted basis, further research is needed into an 8% equivalent MLR and FFS. An insurance model should require PBMs to incur penalties for exceeding contracted delivered trend. The starting point for a MLR of 92% seems high when compared to the 85% MLR mandated by the Affordable Care Act for Medicare Part D plans.[12] As compensation for incurring increased risk, a MLR and FFS equivalent to 10% to 12% seem like a fairer starting point for discussions.

To continue to motivate the Big 3 PBMs to bargain hard with Pharma, and to design-in cost-effective utilization management programs, there needs to be some penalties for not meeting KPIs. At the simplest level, it could be something like the PBM retaining 50% of all trend below contract and absorbing 100% of all trend above contracted PMPY. Again, there needs to be a lot more research around fair trade-offs between contracted FFS and risk-sharing arrangements.

**Changes in PBM Management**

We conclude this paper with some thoughts about likely changes to PBM management of plans underwritten by an insurance business model . Broadly, PBMs will switch to managing to achieve better outcomes as measured by PMPY trend instead of better gross profit margins.

Expect PBMs significantly to increase utilization management programs like prior authorization, step therapy, and quantity limits. Patient advocacy groups will not be pleased. They lobbied for more PBM transparency, and the result will be more utilization management than ever. The danger here is that PBMs will have less incentive to manage drug adherence. There needs to be clauses in PBM insurance contracts specifying KPIs for adherence and patient well-being. There also needs to be a clause in contracts preventing a year-end tsunami of additional utilization restrictions so as not to exceed contract trend..

______________________

12 Center for Medicare and Medicaid Services, *Medical Loss Ratio, 2024,*
https://www.cms.gov/medicare/health-drug-plans/medical-loss-ratio

The gross to net price bubble will shrink even further as only net prices matter, not gross rebates. Generics and low list price biosimilars will now uniformly be favored over higher list price brands that cannot prove superior efficacy. The convoluted PBM scheme of private labeling biosimilars will be scrapped as the scheme's opaque margins no longer matter.

The financials of retail pharmacies likely will not improve under an insurance model. PBMs still will bargain hard over retail pharmacy reimbursements in the desire to reduce the overall trend. The only saving grace for retail pharmacies is that PBMs will care only about prices of Rx regardless of where they are dispensed. Expect PBMs finally to allow retail pharmacies to dispense 90-day maintenance Rx if price competitive.

In terms of the design of rebate contracts, formulary bid menus will be simplified by eliminating opaque administrative fees and price protection rebates. On the other hand, incremental rebates for outright exclusion of named drug competitors will continue and outright formulary exclusions will continue. With the threat of a "world without rebates" diminished, the motivation for offshore rebate aggregators is gone.

Nominally, it is plan sponsors, not PBMs, who set copayment dollars and coinsurance percentages. While not a utilization management technique, they do affect utilization. Operating under an insurance business model, PBM likely would be averse to any ratchet down of copayments and coinsurance unless offset by a higher KPI for delivered trend.

It is also interesting to note here that the vertical integration of the Big 3 PBMs with insurance companies has reduced the legal and actuarial expertise required to launch insurance-based pharmacy benefit plans. It is interesting to speculate which companies will be first to offer these plans.

We think leaders will be companies historically dominated by the insurance business rather than the PBM business. We see United Healthcare and Elevance (formerly Anthem Blue Cross and Blue Shield) as leaders in offering insurance pharmacy benefit plans. In fact, United Healthcare's PBM OptumRx quietly launched in 2025 a so-called "Clear Cut Trend" plan that guarantees a PMPY drug trend.[13] However, we could find no publicly available specifics such as trend guarantees, risk sharing or FFS.

Discussions about moving to a pharmacy benefit insurance model leads to thoughts about coverage for pharmacy and medical drugs under a single medical insurance plan. Many biologics now are available as infusions or self-injectables with the single HCPCS code for the drug itself and two different codes for delivery. The prospects for disease management would seem better if there were no longer two separate plans for specialty drugs.

______________________

13 UnitedHealth Group , *OptumRx Announces New Pricing Model,* May 2024, https://www.unitedhealthgroup.com/newsroom/posts/2024/2024-05-optum-rx-clear-trend-guarantee.

## Disclosures:

I have not received any remuneration for this paper nor financial interest in any company cited in this working paper. I have a Ph.D. in Economics from Washington University in St. Louis and a B.A. in Economics from Amherst College. Other papers on PBMs can be found on my website  https://labrams.co

© Lawrence W. Abrams, 2025